\documentclass[prd,aps,10pt,superscriptaddress,preprintnumbers,notitlepage,nofootinbib,longbibliography]{revtex4-2}
\usepackage{amssymb,amsthm,amsmath}
\usepackage{textcomp}
\usepackage{color}      
\usepackage{slashed}    
\usepackage{verbatim}
\usepackage[normalem]{ulem} 
\usepackage{hyperref}
\usepackage{slashed}   
\usepackage{soul}
\usepackage{xcolor}
\usepackage[compat=1.1.0]{tikz-feynman}
\usepackage{float}
\usepackage{braket}
\usepackage{booktabs}
\usepackage{rotating}  
\usepackage{multirow}  
\usepackage{simpler-wick}
\usepackage{caption}
\usepackage{subcaption}
\usepackage{epsfig}
\newcommand{\hatp}{\hat{\boldsymbol{p}}}
\newcommand{\hatk}{\hat{\boldsymbol{k}}}

\newcommand{\hatl}{\hat{\boldsymbol{l}}}
\usepackage{iftex}
\ifPDFTeX
\DeclareUnicodeCharacter{2212}{-}
\DeclareUnicodeCharacter{03C4}{\ensuremath{\tau}}
\DeclareUnicodeCharacter{202F}{\,}
\fi
\hypersetup{pdftitle={Probing the imaginary parts and their q2 dependences for the tau g-2 and EDM},
 pdfauthor={Xin-Yu Du, Xiao-Gang He, Zhong-Lv Huang, Chia-Wei Liu, Zi-Yue Zou}}
\hypersetup{hidelinks}
\AtBeginDocument{}
\begin{document}

	\title{Probing the imaginary parts and  their $q^2$ dependences for the tau $g-2$ and  EDM}

\author { Xin-Yu Du}\email{2020dxy@sjtu.edu.cn}

	\affiliation{
		State Key Laboratory of Dark Matter Physics,
		Tsung-Dao Lee Institute \& School of Physics and 
		Astronomy, Shanghai Jiao Tong University, Shanghai 200240, China} 
	\affiliation{Key Laboratory for Particle Astrophysics and Cosmology (MOE)
		\& Shanghai Key Laboratory for Particle Physics and Cosmology,
	Tsung-Dao Lee Institute  \&	School of Physics and Astronomy, Shanghai Jiao Tong University, Shanghai 200240, China}
    
    \author {Xiao-Gang He}\email{hexg@sjtu.edu.cn}
			
		\affiliation{
		State Key Laboratory of Dark Matter Physics,
		Tsung-Dao Lee Institute \& School of Physics and 
		Astronomy, Shanghai Jiao Tong University, Shanghai 200240, China} 
	\affiliation{Key Laboratory for Particle Astrophysics and Cosmology (MOE)
		\& Shanghai Key Laboratory for Particle Physics and Cosmology,
		Tsung-Dao Lee Institute  \&	School of Physics and Astronomy, Shanghai Jiao Tong University, Shanghai 200240, China}
        
	\author { Zhong-Lv Huang}\email{huangzhonglv@sjtu.edu.cn}
    
		\affiliation{
		State Key Laboratory of Dark Matter Physics,
		Tsung-Dao Lee Institute \& School of Physics and 
		Astronomy, Shanghai Jiao Tong University, Shanghai 200240, China} 
	\affiliation{Key Laboratory for Particle Astrophysics and Cosmology (MOE)
		\& Shanghai Key Laboratory for Particle Physics and Cosmology,
		Tsung-Dao Lee Institute  \&	School of Physics and Astronomy, Shanghai Jiao Tong University, Shanghai 200240, China}
	
	  \author {Chia-Wei Liu}\email{chiaweiliu@ucas.ac.cn}
    \affiliation{School of Fundamental Physics and Mathematical Sciences, Hangzhou Institute for Advanced Study, UCAS, Hangzhou 310024, China}
    \author {Zi-Yue Zou}\email{ziy\_zou@sjtu.edu.cn}
			
		\affiliation{
		State Key Laboratory of Dark Matter Physics,
		Tsung-Dao Lee Institute \& School of Physics and 
		Astronomy, Shanghai Jiao Tong University, Shanghai 200240, China} 
	\affiliation{Key Laboratory for Particle Astrophysics and Cosmology (MOE)
		\& Shanghai Key Laboratory for Particle Physics and Cosmology,
		Tsung-Dao Lee Institute  \&	School of Physics and Astronomy, Shanghai Jiao Tong University, Shanghai 200240, China}

	\date{\today}

    \begin{abstract}
    The $\tau$ anomalous magnetic dipole moment (MDM) $a_\tau = (g-2)_\tau/2$ and electric dipole moment (EDM) $d_\tau$, are precision probes of electroweak dynamics and possible new physics sources, yet both remain weakly constrained experimentally. Treated as generalized form factors, these quantities exhibit a generic $q^2$ dependence for an off-shell interacting photon. For timelike momentum transfer above the $\tau^+\tau^-$ threshold, $q^2 = s > 4m_\tau^2$, the form factors can acquire absorptive imaginary parts. We investigate how such a $q^2$ dependence and the associated imaginary parts are generated from two distinct perspectives: a heavy-scale effective-theory illustration and an explicit light-scalar Two-Higgs-Doublet Model (2HDM). The effective framework reveals the intimate correlation between $a_\tau$ and $d_\tau$. New CP-violating interactions which generate a non-zero $d_{\tau}$, can also generically have non-zero contributions to $a_\tau$, thereby deeply linking their phenomenological studies. Within the 2HDM, we include precision-decay constraints and identify a benchmark near the $3\sigma$ $Z$-pole compatibility boundary with an imaginary MDM at the projected Belle~II statistical sensitivity. Motivated by these features, we propose experimental methods to extract both the real and imaginary components of the dipole form factors. At the statistical level, these techniques offer Belle~II and the Super Tau-Charm Facility (STCF) the prospect of improving existing bounds on $a_\tau$ by more than one order of magnitude. Finally, we highlight that combining measurements across the distinct center-of-mass energies of Belle II and STCF provides a unique, previously unexplored avenue to explicitly obtain information about the $q^2$ evolution of these dipole form factors.
\end{abstract}
    
\maketitle
\section{Introduction}

The anomalous magnetic dipole moment (MDM), $a_f = (g-2)_f/2$, and the electric dipole moment (EDM), $d_f$ are among the most fundamental quantities of elementary fermions. In the Standard Model (SM), charged-lepton dipole moments receive nonzero loop-level contributions from electroweak interactions, with hadronic effects entering through higher-order corrections~\cite{Aoyama:2020ynm, Eidelman:2007sb, Jegerlehner:2009ry,Schwinger:1948iu,Yamaguchi:2020eub,Yamaguchi:2020dsy,Chupp:2017rkp,Hoogeveen:1990cb,Bernreuther:1990jx}. More generally, they are defined as form factors in the electromagnetic current. When the photon momentum $q$ is off-shell ($q^2 \neq 0$), they naturally develop a $q^2$ dependence. Notably, in the timelike kinematic region ($q^2 = s > 4m_f^2$), both $a_ f(q^2)$ and $d_f(q^2)$ can develop non-vanishing imaginary absorptive parts~\cite{Cutkosky:1960sp, Eden:1966dnq}. 

While the electron and muon dipole moments have been tested with very high precision, the corresponding constraints on the $\tau$ anomalous MDM remain much weaker. According to the Particle Data Group (PDG)~\cite{ParticleDataGroup:2024cfk}, which adopts the result from ATLAS~\cite{ATLAS:2022ryk}, the current allowed interval for \(a_\tau\) is
\begin{equation}
     -0.057 < a_\tau < 0.024, \quad \text{C.L.} = 95 \% .
     \label{Atlas}
\end{equation}

This precision is still far from that required to test the QED prediction at one loop level, $a_\tau^{\text{SM}} \simeq \alpha_{em}/(2\pi) \approx 1.1614 \times 10^{-3}$~\cite{Schwinger:1948iu, Eidelman:2007sb, Aoyama:2020ynm}. The difficulty to have higher precision for $a_\tau$ primarily stems from the extremely short lifetime of the $\tau$ lepton and missing energy in its decays due to production of $\tau$ neutrino~\cite{ParticleDataGroup:2024cfk,Bernreuther:1996dr,Huang:1996jr,Bernreuther:2021elu,Sun:2024vcd,He:2025ewk,Lu:2025heu,Nakai:2025dmp,Fael:2013ij}. Consequently, there is a pressing need for novel experimental approaches to test the SM in the $\tau$ sector at the one-loop level or better, and to probe potential new physics (NP) contributions beyond the SM. Many efforts have been done so far, both experimentally and theoretically~\cite{ ATLAS:2022ryk, CMS:2022arf, Lu:2025heu,Crivellin:2021spu,Hoferichter:2025ijh,Hoferichter:2025zjp,Gogniat:2026zvf}.

Regarding the EDM which violates CP symmetry, the SM prediction is highly suppressed~\cite{Chupp:2017rkp,Hoogeveen:1990cb,Bernreuther:1990jx,Yamaguchi:2020eub,Yamaguchi:2020dsy}, lying many orders of magnitude below current experimental sensitivities. This strong suppression makes the EDM a particularly clean probe of new CP-violating interactions~\cite{Barr:1990vd,Shu:2013uua,Jung:2013hka,Inoue:2014nva,He:1992dc,Dorsner:2016wpm,Fuyuto:2018scm,Dekens:2018bci,Ramsey-Musolf:2006evg,Li:2010ax,Li:2021xmw}. The same as the anomalous MDM, the $\tau$ EDM is much less constrained than the electron and muon EDMs. 
The Belle collaboration has obtained  the following best constraints on the real and imaginary parts of the $\tau$ EDM~\cite{Belle:2021ybo}
\begin{equation}
         \mathrm{Re}(d_{\tau})   =  (-6.2\pm6.3)\times10^{-18}~e\,\mathrm{cm},\;\;\;\;\mathrm{Im}(d_{\tau})    =  (-4.0\pm3.2)\times10^{-18}~e\,\mathrm{cm}.
     \label{Belle}
 \end{equation} 
If NP introduces additional sources of CP violation, $d_\tau$ can be enhanced to a level accessible within the reach of near-future experiments, such as Belle II and the Super Tau-Charm Facility (STCF). The same interactions generating $d_\tau$ may also modify $a_\tau$, making a combined analysis of $a_\tau$ and $d_\tau$ well motivated. To compare the current sensitivities to $a_\tau$ and $d_\tau$ on the same footing, it is instructive to define a counter part $a_\tau$ for $\tau$ EDM, $\tilde{d}_\tau = (2m_\tau/e) d_\tau$. Using the current upper limits provided by the Belle collaboration~\cite{Belle:2021ybo}, the corresponding constraints are 
\begin{eqnarray}
    \mathrm{Re}(\tilde{d}_\tau) = (-1.12\pm1.14)\times10^{-3},\;\;\;\;\mathrm{Im}(\tilde{d}_\tau) = (-7.20\pm5.76)\times10^{-4}.
\end{eqnarray}
This comparison shows that current measurements constrain the dimensionless $\tau$ EDM parameter $\tilde d_\tau$  more strongly than $a_\tau$. Future measurements should therefore aim to improve the sensitivity of $a_\tau$ toward the level required to test the SM loop prediction.

The electromagnetic current vertex for $\gamma \to \tau^+\tau^-$ can be parameterized in terms of form factors as
\begin{equation}
    \Gamma^\mu = F_1(q^2) \gamma^\mu + F_2(q^2) \frac{i \sigma^{\mu\nu} q_\nu}{2 m_\tau} + F_3 (q^2) \sigma^{\mu\nu} q_\nu \gamma_5 + F_4(q^2) (q^2 \gamma^\mu - q^\mu \slashed{q}) \gamma_5 \,.
\label{eq:formfactor}
\end{equation}
Here, the form factors $F_2(q^2)$ and $F_3(q^2)$ are directly related to the generalized anomalous MDM and EDM via $a_{\tau}(q^2)=F_2(q^2)$ and $d_\tau(q^2) = -e F_3(q^2)$, respectively. The conventional $g-2$ and EDM are defined strictly at the photon on-shell limit, $a_\tau \equiv a_\tau(0)$ and $d_\tau \equiv d_\tau(0)$. At STCF and Belle II energies, the accessible quantities are inherently $F_{2}(s)$ and $F_{3}(s)$. In the context of $e^+e^- \to \tau^+\tau^-$ production at these facilities, the momentum transfer squared $q^2$ is identically the center-of-mass energy squared $s$. Thus, the notations $q^2$ and $s$ are used interchangeably hereafter.

\begin{figure}[htbp]
    \centering
    \includegraphics[width=0.4\textwidth]{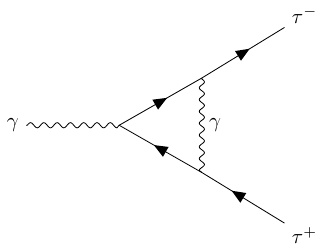}
    
    \caption{One-loop QED vertex correction diagram in the timelike region}
    \label{QED1loop}
\end{figure}
The $F_2$ and $F_3$ structures correspond to dipole operators which are dimensin-5 and are not allowed at elementary tree-level interactions in renormalizable UV-complete theories. But they can be dynamically generated via loop corrections. For instance, in standard QED, we can calculate the one-loop vertex correction in the time like region shown in Fig.~\ref{QED1loop}, starting from the tree level interaction with $F_1=1$. One obtains
\begin{align}
    &a_\tau^{\text{QED}_{1\text{-loop}}}(s) = \frac{\alpha_{\text{em}} m_\tau^2}{\pi s \beta_\tau} \ln\left( \frac{\beta_\tau-1}{1+\beta_\tau} \right), \label{eq:atau_qed}  \\
    &\text{Re}\big[a_\tau^{\text{QED}_{1\text{-loop}}}(s)\big] = \frac{\alpha_{\text{em}} m_\tau^2}{\pi s \beta_\tau} \ln\left( \frac{1-\beta_\tau}{1+\beta_\tau} \right), \label{eq:atau_1loop} \\
    &\text{Im}\big[a_\tau^{\text{QED}_{1\text{-loop}}}(s)\big] = \frac{\alpha_{\text{em}} m_\tau^2}{s \beta_\tau}\Theta(s-4m_\tau^2).\label{eq:atau_im_1loop} 
\end{align}
where $\alpha_{\text{em}} = e^2/(4\pi)$ is the fine-structure constant, and $\beta_\tau \equiv \sqrt{1 - 4m_\tau^2/s}$ is the velocity of the $\tau$ lepton. Analytically continuing this dispersive expression to the static limit ($s \to 0$) exactly recovers the classic Schwinger result, $a_\tau(0) = \alpha_{\text{em}}/(2\pi)$. Since pure QED preserves CP symmetry, no $\tau$ EDM is generated at this order. According to the Cutkosky cutting rules, the imaginary part arises when $s$ exceeds the kinematic threshold ($s > 4m_\tau^2$), allowing the intermediate loop $\tau$ leptons to go on-shell, as implied by Eq.~\eqref{eq:atau_im_1loop}. 

In the SM, $d_\tau$ is generated at the four-loop level and is quite small. EDMs are
	generated only through higher-order loop processes and therefore are predicted to be extremely small~\cite{Hoogeveen:1990cb,Bernreuther:1990jx,Chupp:2017rkp,Yamaguchi:2020eub,Yamaguchi:2020dsy}.
Studying the behavior of its imaginary part and $q^2$ dependence is highly involved. However, in scenarios beyond the SM, these properties may appear at the one-loop level and can be studied in a more straightforward manner, which will be one subject of this paper.

In this work, we investigate $a_\tau$ and $d_\tau$ from two complementary theoretical perspectives. First, we utilize the Standard Model Effective Field Theory (SMEFT). Within this framework, $a_\tau(s)$ and $d_\tau(s)$ receive contributions from two types of operators: dipole operators via direct tree-level insertions, and four-fermion operators via one-loop $\tau$ diagrams. We evaluate the latter process to explicitly demonstrate the generation of the imaginary part and the non-trivial $q^2$ dependence. Furthermore, this approach illustrates how $a_\tau$ and $d_\tau$ can be correlated by common effective operators.

Second, we study a separate light-mediator example in a renormalizable 2HDM, retaining the scalar propagators rather than matching the light state onto local four-fermion operators. Precision $Z$ and tau decays restrict the allowed Yukawa size; we use a benchmark near the adopted $3\sigma$ $Z$ boundary to illustrate the remaining statistical reach.

Finally, recognizing the necessity for enhanced experimental precision to probe these theoretical predictions, we study novel experimental methodologies to measure the dipole form factors at $e^+e^-$ colliders. By systematically exploiting detailed angular distributions and spin correlations in $\tau^+\tau^-$ production and decay, we find that the current bounds on both the real and imaginary parts of $a_\tau$ and $d_\tau$ can be improved considerably.

\section{EFT generation of the Timelike $\tau$ anomalous MDM and EDM }\label{sec:eft}

To explicitly demonstrate the physical significance of the $q^2$ dependence and the imaginary parts of the $\tau$ anomalous MDM and EDM form factors, we perform a model-independent analysis within the framework of Effective Field Theory (EFT). We use representative four-fermion and dipole operators, not an exhaustive operator classification. A controlled LEFT expansion requires $\sqrt{s}\ll\Lambda$, while its SMEFT embedding below additionally assumes $\Lambda\gg v$. This section is not a low-energy expansion of the light-$A$ benchmark in Sec.~\ref{sec:uvmodel}. 

At energy scales well below the electroweak scale, the relevant NP interactions can be parameterized in the Low-Energy Effective Field Theory (LEFT). We focus on a representative set of leading operators, consisting of four-fermion scalar-like interactions and direct dipole interactions~\cite{Jenkins:2017jig}
\begin{equation}
    {\cal L}_{\mathrm{eff}} \supset 
    \frac{C_{SS}^{\tau\tau}}{\Lambda^2}(\bar\tau\tau)^2
    + \frac{C_{PP}^{\tau\tau}}{\Lambda^2}(\bar\tau i\gamma_5\tau)^2 
    + \frac{ C_{SP}^{\tau \tau } }{\Lambda^2 }  (\overline{\tau} i\gamma_5 \tau ) (\overline{\tau} \tau )
    - \frac{e a_{\tau}^0}{4m_{\tau}}F^{\mu\nu} \overline{\tau } \sigma _{\mu \nu} \tau 
    - d_\tau^0 \frac{i}{2} F^{\mu\nu} \overline{\tau } \sigma _{\mu \nu} \gamma_5 \tau \,.
\end{equation}
To ensure the Hermiticity of the Lagrangian, all parameters ($C_{SS}^{\tau\tau}, C_{PP}^{\tau\tau}, C_{SP}^{\tau \tau}, a_\tau^0$, and $d_\tau^0$) must be strictly real. Consequently, at tree level, these operators do not   generate any imaginary parts. Based on their CP properties, the CP-even coefficients ($C_{SS}^{\tau\tau}, C_{PP}^{\tau\tau}, a_\tau^0$) contribute to the anomalous MDM, while the CP-odd coefficients ($C_{SP}^{\tau\tau}, d_\tau^0$) contribute to the EDM. 

If the NP scale $\Lambda$ is much higher than the electroweak scale, these LEFT operators should be matched onto operators invariant under the full SM gauge group~\cite{Grzadkowski:2010es,Murphy:2020dim8}, $SU(3)_C \times SU(2)_L \times U(1)_Y$
\begin{equation}
    \mathcal{L}_{\rm SMEFT} \supset
    \frac{C^{33}_{(eH)^2}}{\Lambda^{4}} \big[(\overline{L}_\tau H)\tau_R\big]^2
    + \frac{C^{33}_{eB}}{\Lambda^{2}} (\overline{L}_\tau \sigma^{\mu\nu}\tau_R) H B_{\mu\nu}
    + \frac{C^{33}_{eW}}{\Lambda^{2}} (\overline{L}_\tau \sigma_{\mu\nu}\tau_R) \sigma_a H W_a^{\mu\nu}
    + \text{h.c.} \,.
\end{equation} 
where $H$ denotes the Higgs doublet, $B$ and $W$ are the SM gauge fields, and $\sigma_a$ are the Pauli matrix. For the same-flavor chirality structure displayed here, the CP-odd scalar four-tau interaction has a dimension-eight SMEFT completion. After the spontaneous symmetry breaking, the relevant SMEFT Wilson coefficients are matched onto the LEFT parameters as~\cite{Grzadkowski:2010es,Alonso:2013hga}
\begin{equation}
    \begin{aligned}
        C_{SS}^{\tau\tau} &= -C_{PP}^{\tau\tau} = \frac{v^2}{4\Lambda^{2}}\mathrm{Re}[C_{(eH)^2}^{33}], \quad
        &C_{SP}^{\tau \tau} &= \frac{v^{2}}{2\Lambda^{2}} \mathrm{Im}\!\big[C^{33}_{(eH)^2}\big], \\
        a_\tau^0 &= \frac{-2\sqrt2\,v\,m_\tau}{e\Lambda^2}\, \mathrm{Re}\!\big[c_W C_{eB}^{33}-s_W C_{eW}^{33}\big], \quad
        &d_\tau ^0 &= -\frac{\sqrt{2} v}{\Lambda^{2}} \mathrm{Im}\!\Big[c_W\,C^{33}_{eB}-s_W\,C^{33}_{eW}\Big] .
    \end{aligned}
\end{equation}
Here, $v/\sqrt{2}$ is the vacuum expectation value of $H$, and $(c_W,s_W) = ( \cos \theta_W , \sin \theta_W )$. This matching shows that the anomalous MDM and EDM are intimately correlated.

\begin{figure}[htbp]
    
        \centering
         \includegraphics[width=0.48\textwidth]{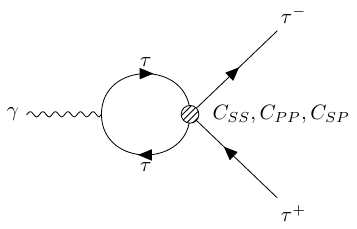}
        \caption{Illustrative Feynman diagrams of the one-loop dynamical processes generating the absorptive imaginary parts of the effective form factors using four-fermion operators ($C_{SP, SS}$) via a closed $\tau$ loop, which yields the kinematic function $g(s)$.}
    \label{fig:eft_diagrams}
\end{figure}
To illustrate the dynamical generation of the effective form factors, denoted as $a^{{\mathrm{EFF}}}_\tau(s)$ 
and $d^{{\mathrm{EFF}}}_\tau(s)$, we evaluate the contributions from two distinct processes. In this illustrative calculation, we assume the standard QED vector coupling is not significantly modified by NP. First, the tree-level insertion of the dipole operators directly yields the constants $a_\tau^0$ and $d_\tau^0$. Due to Hermiticity, this process contributes exclusively to the real parts and generates no $q^2$ dependence or imaginary components. Second, we consider the one-loop diagram involving a four-fermion operator and a standard QED vertex, forming a closed $\tau$ loop as shown in Fig.~\ref{fig:eft_diagrams}. In the timelike region ($s > 4m_\tau^2$), this diagram generate an absorptive imaginary part parameterized by a function $g(s)$ given below.
 
Combining the two contributions discussed above, the effective form factors are given by
\begin{align}
    a^{\mathrm{EFF}}_\tau(s) &= a_\tau^0  + \frac{ m_\tau^2 (C_{SS}^{\tau\tau}-C_{PP}^{\tau\tau})}{4\pi^2\Lambda^2} g(s) \,, \label{eq:a_tau} \\
    d^{\mathrm{EFF}}_\tau(s) &= d_\tau^0 + \frac{e m_\tau C_{\text{SP}}^{\tau\tau}}{8\pi^2\Lambda^2} g(s) \,. \label{eq:d_tau}
\end{align}
where the kinematic loop functions are analytically derived as
\begin{align}
    g(s) &= 2 + \beta_\tau \ln\left( 1 - \frac{s}{2m_\tau^2}(1-\beta_\tau) \right). 
\end{align}
The form factors are evaluated at the renormalization scale $\mu = m_\tau$ in the $\overline{\text{MS}}$ scheme. Crucially, the absorptive imaginary parts of these form factors depend on the $q^2=s$ kinematics
\begin{align}
    \text{Im}(a^{\mathrm{EFF}}_\tau(s)) &= \frac{m_\tau^2 (C_{SS}^{\tau \tau }-C_{PP}^{\tau \tau }) }{4 \pi \Lambda^2}\beta_\tau \Theta(s-4m_\tau^2)  ,\label{eq:Im_a_eff} \\
    \text{Im}(d^{\mathrm{EFF}}_\tau(s)) &= \frac{e m_\tau C_{SP}^{\tau \tau } }{8 \pi \Lambda^2}\beta_\tau \Theta(s-4m_\tau^2) \, 
    .\label{eq:Im_d_eff}
\end{align}

The absorptive parts in Eqs.~\eqref{eq:Im_a_eff} and \eqref{eq:Im_d_eff} depend on the kinematics and four-tau coefficients, whereas a Hermitian local dipole coefficient is real. Indirect electron-EDM limits constrain a pointlike tau dipole under specified matching assumptions~\cite{Laporta:1992pa,Grozin:2008nw}. With only this dipole source and no cancellations, the $90\%$-confidence bound $|d_e|\leq4.1\times10^{-30}\,e\,\mathrm{cm}$~\cite{Roussy:2022cmp} gives~\cite{Ema:2022wxd}
\begin{equation}\label{eq:tradi}
    | d_\tau^0 | \leq 4.1 \times 10^{-19}~e\,\mathrm{cm} \,. 
\end{equation}
This is not a model-independent constraint on each contribution to a momentum-dependent form factor. Four-fermion operators can feed electron EDMs through running and matching, and are not generically immune to such constraints. A joint matching analysis is beyond the present illustrative calculation. For a sufficiently small local $d_\tau^0$, the timelike form factor may nevertheless be dominated by the four-tau term. The Belle intervals~\cite{Belle:2021ybo} then give the formal coefficient relations
\begin{align}
    \left|\text{Re}\left( d_\tau\big((10.58~\text{GeV})^2\big) \right)\right| &= \left| d_\tau^0- 1.3 \frac{e m_\tau C_{SP}^{\tau \tau }}{ 8 \pi^2 \Lambda^2 } \right| \leq 1.66 \times 10^{-17}~e\,\mathrm{cm} \,, \\
    \left|\text{Im}\left( d_\tau\big((10.58~\text{GeV})^2\big) \right)\right| &= \left| 3 \frac{e m_\tau C_{SP}^{\tau \tau }}{ 8 \pi^2 \Lambda^2 } \right| \leq 0.93 \times 10^{-17}~e\,\mathrm{cm} \,.
\end{align}
These inequalities constrain $C_{SP}^{\tau\tau}/\Lambda^2$, not a measured cutoff. Saturating them with order-one coefficients at $\Lambda$ comparable to $\sqrt{s}$ would lie outside the EFT validity domain; we use no such saturated EFT benchmark. At $\sqrt{s}=6.3$~GeV~\cite{He:2025ewk}, the corresponding coefficient dependence is
\begin{align}
    \left|\text{Re}\left( d_\tau\big((6.3~\text{GeV})^2\big) \right)\right| &= \left| d_\tau^0+  0.06 \frac{e m_\tau C_{SP}^{\tau \tau }}{ 8 \pi^2 \Lambda^2 } \right| \,, \\
    \left|\text{Im}\left(d_\tau\big((6.3~\text{GeV})^2\big) \right)\right| &= \left| 2.6 \frac{e m_\tau C_{SP}^{\tau \tau }}{ 8 \pi^2 \Lambda^2 } \right| \,. 
\end{align}
It is worth noting that the dynamic emergence of form factor imaginary parts from effective operators has also been investigated in recent literature. Specifically, Ref.~\cite{Gogniat:2026zvf} evaluated the subleading electroweak background and the chirality-suppressed four-fermion interference at the loop level for $\tau$ dipole measurements at Belle II. Furthermore, Refs.~\cite{Hoferichter:2025ijh,Hoferichter:2025zjp} demonstrated that for light New Physics scenarios, the absorptive parts generated by loop-level insertions can induce non-vanishing experimental asymmetries. Crucially, these studies show that such imaginary parts can be extracted using currently available unpolarized data at Belle II, without relying on future beam polarization upgrades~\cite{Crivellin:2021spu}. 
Our calculation illustrates the distinct, $q^2$-dependent absorptive signatures of four-tau interactions; it does not establish their independence from all low-energy EDM constraints.

Having illustrated these effects in the heavy-scale EFT, we next calculate them directly in a separate renormalizable light-mediator model, without expanding in $s/m_A^2$.

\section{A UV-complete model study and Phenomenological Analysis}\label{sec:uvmodel}

We now study a UV-complete realization that can generate sizable contributions to $a_\tau$ and $d_\tau$. We begin with a generic CP-violating interaction between a neutral scalar and the $\tau$ lepton. In our previous work~\cite{Huang:2025ghw}, we studied the contribution of such an interaction to $\tau$ EDM. A natural physical extension is that these identical dynamical processes will also contribute to $a_\tau$. As explicitly demonstrated in our previous study, the EDM form factor can be enhanced when the scalar is light and becomes suppressed for larger scalar masses. We find that a similar mass dependence occurs for the anomalous MDM. Therefore, we focus on a new scalar particle in the GeV-scale mass range. 

In realistic UV-complete setups like 2HDM, which can contain several neutral scalars, the dominant contribution to the dipole form factors can arise from the lightest neutral state.
We parameterize the interaction of one neutral spin-0 particle, $A$, with the $\tau$ lepton as
\begin{equation}
    \mathcal{L}_A =  A \bar{\tau} (a - i b \gamma_5) \tau \,. \label{eq:Att-int}
\end{equation}
where $a$ and $b$ are the real scalar (CP-even) and pseudoscalar (CP-odd) coupling constants, respectively. Using this interaction Lagrangian, the one-loop NP contribution to the $\tau$ anomalous MDM form factor can be decomposed into a scalar component and a pseudoscalar component
\begin{equation}
    a_{\tau}^{\text{NP}}(s) = a^2 f_1(s, m_A) + b^2 f_2(s, m_A) \,.
    \label{eq:mdm_decoupled}
\end{equation}
where $f_1$ and $f_2$ are given by 
\begin{widetext}
    \begin{equation}\label{eq:amdm_full}
        \begin{split}
            f_1(s, m_A) &= \frac{1}{16\pi^2 (s - 4m_\tau^2)^2} \Bigg\{ 2m_A^2(s - 4m_\tau^2) + 2\big( 16m_\tau^4 + m_A^2 s - 2m_\tau^2(5m_A^2 + 2s) \big) B(m_\tau^2, m_\tau, m_A) \\
            &\quad + 6m_\tau^2 (s - 4m_\tau^2 + 2m_A^2) B(s, m_\tau, m_\tau) \\
            &\quad + \frac{m_A^2}{m_\tau^2} \big( 24m_\tau^4 + m_A^2 s - 2m_\tau^2(5m_A^2 + 3s) \big) \ln\left(\frac{m_\tau^2}{m_A^2}\right) \\
            &\quad + 12m_\tau^2 m_A^2 (s - 4m_\tau^2 + m_A^2) C_0(m_\tau^2, m_\tau^2, s, m_\tau, m_A, m_\tau) \Bigg\}, \\
            f_2(s, m_A) &= \frac{1}{16\pi^2 (s - 4m_\tau^2)^2} \Bigg\{ 2m_A^2(s - 4m_\tau^2) - 2m_A^2(10m_\tau^2 - s) B(m_\tau^2, m_\tau, m_A) \\
            &\quad - 2m_\tau^2 (s - 4m_\tau^2 - 6m_A^2) B(s, m_\tau, m_\tau) \\
            &\quad + \frac{m_A^2}{m_\tau^2} \big( 8m_\tau^4 + m_A^2 s - 2m_\tau^2(5m_A^2 + s) \big) \ln\left(\frac{m_\tau^2}{m_A^2}\right) \\
            &\quad + 4m_\tau^2 m_A^2 (s - 4m_\tau^2 + 3m_A^2) C_0(m_\tau^2, m_\tau^2, s, m_\tau, m_A, m_\tau) \Bigg\} \,.
        \end{split}
    \end{equation}
\end{widetext}
where
\begin{equation}
	B(s, m_0, m_1) = \frac{\sqrt{\lambda(s, m_0^2, m_1^2)}}
	{s} \ln\left( \frac{m_0^2 + m_1^2 - s + \sqrt{\lambda(s, m_0^2, m_1^2)}}{2 m_0 m_1} \right) \,.
	\label{eq:DiscB}
\end{equation}
with $\lambda(x, y, z) = x^2 + y^2 + z^2 - 2xy - 2yz - 2zx$, and
\begin{align}
	C_0(m_0^2, m_0^2, q ^2, m_0 , m_1 , m_0 ) = -\int_0^1 dx_1 \int_0^{1-x_1} dx_2 \frac{1}{ 
		(1 - x_1)^2 m_0^2
		\;+\;
		x_1 m_1^2
		\;+\;
		x_2 (x_1 + x_2 - 1)\, q ^2
	}.
\end{align}
The CP-violating EDM form factor can be written by
\begin{eqnarray}\label{eq:EDM-tau_full}
    \begin{aligned}
        d^{\text{NP}}_\tau (s) &= \frac{e ab}{4 \pi^2 m_\tau (s - 4 m_\tau^2) } \Big[ m_\tau^2 B(m_\tau^2, m_\tau ,m_A) - m_\tau^2 B(s, m_\tau, m_\tau) \\
        &+ m_A^2 \left( \ln\left(\frac{m_\tau}{m_A}\right) - m_\tau^2 C_0(m_\tau^2, m_\tau^2, s, m_\tau, m_A, m_\tau) \right) \Big] \,.
    \end{aligned}
\end{eqnarray}

\begin{figure}[htbp]
    \centering
    \includegraphics[width=0.48\textwidth]{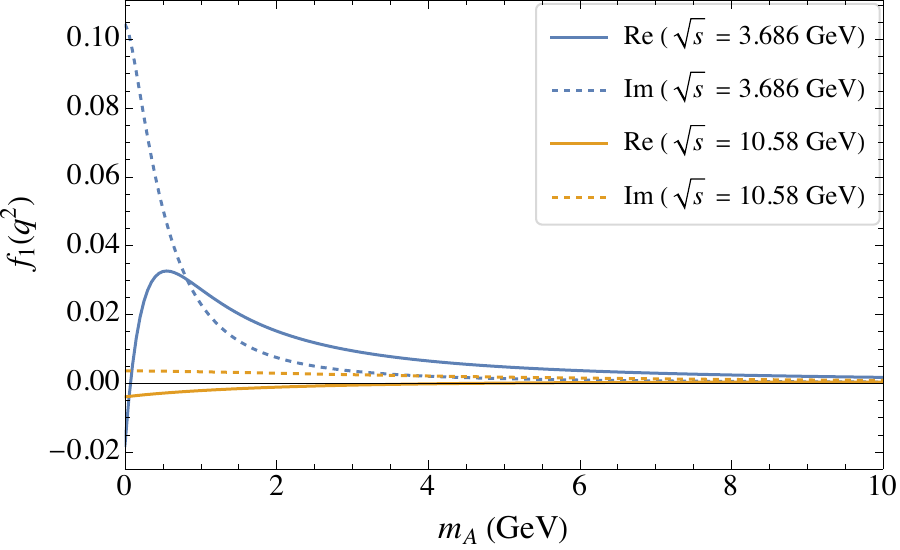}
    \hfill
    \includegraphics[width=0.48\textwidth]{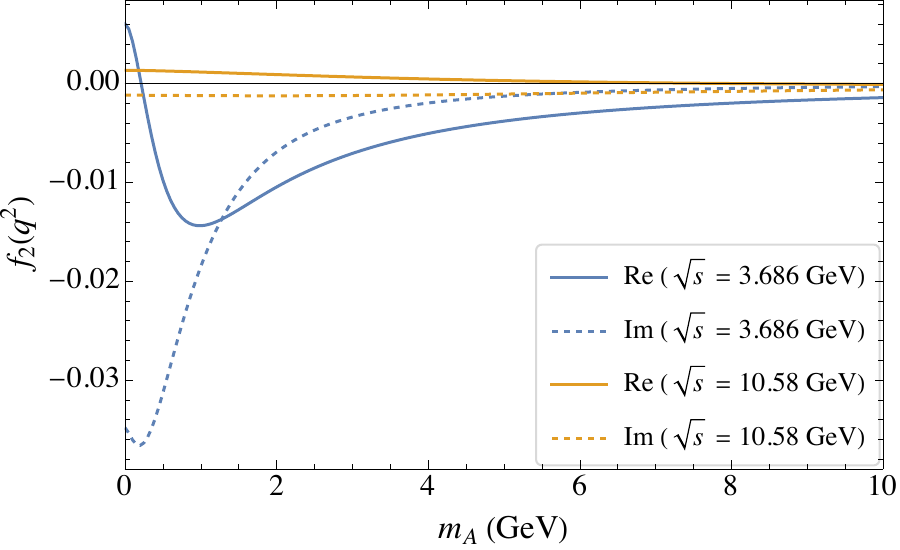}
    \caption{The real (solid lines) and imaginary (dashed lines) parts of the scalar loop function $f_1(s,m_A)$ (left) and the pseudoscalar loop function $f_2(s,m_A)$ (right) as functions of the scalar mass $m_A$. The blue and orange curves correspond to the center-of-mass energies of STCF ($\sqrt{s} = 3.686$~GeV) and Belle~II ($\sqrt{s} = 10.58$~GeV), respectively. The real parts of $f_1$ and $f_2$  have opposite signs over much of the light-mass region, leading to destructive interference in the anomalous MDM when a and b are comparable.}
    \label{fig:f1f2_plots}
\end{figure}
One can see that a sizable anomalous MDM is predominantly generated when the scalar mediator is light from Fig.~\ref{fig:f1f2_plots}, while being heavily suppressed at larger masses. In a UV-complete setup such as the 2HDM, the lightest neutral scalar can dominate the dipole phenomenology when the heavier states are sufficiently decoupled.
The scalar and pseudoscalar terms in Eq.~\eqref{eq:amdm_full} can partially cancel, as shown in Fig.~\ref{fig:f1f2_plots}. We retain a scalar-dominated benchmark to avoid this cancellation; the Higgs-decay comparison below does not require such a hierarchy.
 The most general scalar potential
 in the 2HDM
 is given by~\cite{Wu:1994ja}:
\begin{eqnarray}
    \begin{aligned}
    V &= \frac{1}{2}\left[m_{11}^{2}|\Phi_{1}|^{2}+m_{22}^{2}|\Phi_{2}|^{2}-\left(m_{12}^{2}\Phi_{1}^{\dagger}\Phi_{2}+\mathrm{h.c.}\right)\right] +\frac{\lambda_{1}}{2}(\Phi_{1}^{\dagger}\Phi_{1})^{2}+\frac{\lambda_{2}}{2}(\Phi_{2}^{\dagger}\Phi_{2})^{2} \\
    &+\lambda_{3}(\Phi_{1}^{\dagger}\Phi_{1})(\Phi_{2}^{\dagger}\Phi_{2})+\lambda_{4}(\Phi_{1}^{\dagger}\Phi_{2})(\Phi_{2}^{\dagger}\Phi_{1}) +\left[\frac{\lambda_{5}}{2}(\Phi_{1}^{\dagger}\Phi_{2})^{2}+\left(\lambda_6\Phi_1^\dagger\Phi_1+\lambda_7\Phi_2^\dagger\Phi_2\right)\left(\Phi_1^\dagger\Phi_2\right)+\mathrm{h.c.}\right].
    \end{aligned}
    \label{eq:general_potential}
\end{eqnarray}
Details of notation are given in Appendix~\ref{2HDM}.

In the following, we work in the Higgs basis, where only one Higgs doublet acquires a nonzero vacuum expectation value. $H$ and $H^\pm$ denote the heavy neutral and charged scalars, $h$ is the SM-like Higgs boson, and the light state $A$ mediates the dipole contributions studied here. In the Higgs basis, and in the alignment limit which can be found in Appendix~\ref{2HDM}, the scalar masses after electroweak symmetry breaking are
\begin{align}
    m^2_h &= \Lambda_1 v^2 \,, & m^2_{H^\pm} &= M^2_{22} + \frac{1}{2}\Lambda_3 v^2 \,, \nonumber \\
    m^2_H &= m^2_{H^\pm} + \frac{1}{2} (\Lambda_4 + \Lambda_5) v^2 \,, & m^2_A &= m^2_{H^\pm} + \frac{1}{2} (\Lambda_4 - \Lambda_5) v^2 \,.
    \label{eq:hmass}
\end{align}
Eq.~(\ref{eq:hmass}) makes explicit how a relatively light neutral particle $A$ can be obtained via an appropriate choice of $(\Lambda_4 - \Lambda_5)$, without compromising the required heaviness of $H$ and $H^\pm$. The Yukawa interaction terms for the neutral particle $A$ and $H$ are
\begin{eqnarray}
	\begin{aligned}
		-\mathcal{L}_Y =
		\bar l \hat M_l l + \bar l 
		\left(\frac{\mathrm{Re}(Y_2)}{\sqrt{2}} + i \frac{\mathrm{Im}(Y_2)}{\sqrt{2}} \gamma_5 \right)
		l H
		+
		\bar l 
		\left(-\frac{\mathrm{Im}(Y_2)}{\sqrt{2}} + i \frac{\mathrm{Re}(Y_2)}{\sqrt{2}} \gamma_5 \right)
		l A\; .
	\end{aligned}{\label{eq:HA-coupling}}
\end{eqnarray}
The general Yukawa interactions mapping to the charged leptons yield effective couplings for $A$
\begin{eqnarray}
    a = - \frac{\mathrm{Im}(Y_2)_{33}}{\sqrt{2}} \;,\quad b = - \frac{\mathrm{Re}(Y_2)_{33}}{\sqrt{2}} \,.
\end{eqnarray}

We impose the tau-specific texture $(Y_2)_{ij}=(Y_2)_{33}\delta_{i3}\delta_{j3}$ in the charged-lepton mass basis, with no additional quark Yukawas. We take a real scalar potential and $\Lambda_6=\Lambda_7=0$ at the reference scale; the complex Yukawa coupling supplies CP violation. Only the light-$A$ contribution is retained in the dipole predictions, whereas the precision-decay corrections include the extra $H,A,H^\pm$ states. The scalar-potential conventions are given in Appendix~\ref{2HDM}.

{
Here we use $m_A = 2\,{\rm GeV},m_H = m_{H^\pm} = 500\,{\rm GeV}$ as input for illustration.
}
Alignment gives SM-like tree-level $hVV$ couplings, consistent with the comparison in Ref.~\cite{CMS:2022uhn}, while $m_H=m_{H^\pm}$ cancels the leading contribution to the electroweak $T$ parameter~\cite{Haber:2010bw}. We also choose the renormalized $hAA$ trilinear coupling to vanish at the reference scale, suppressing the tree-level exotic Higgs width. The adopted charged-scalar mass is well above the LEP mass range excluded in the conventional decay benchmarks~\cite{LEPHiggsWorkingGroupforHiggsbosonsearches:2001ogs,ALEPH:2013htx}.

\begin{figure}[htbp]
    \centering
    \includegraphics[width=0.56\textwidth]{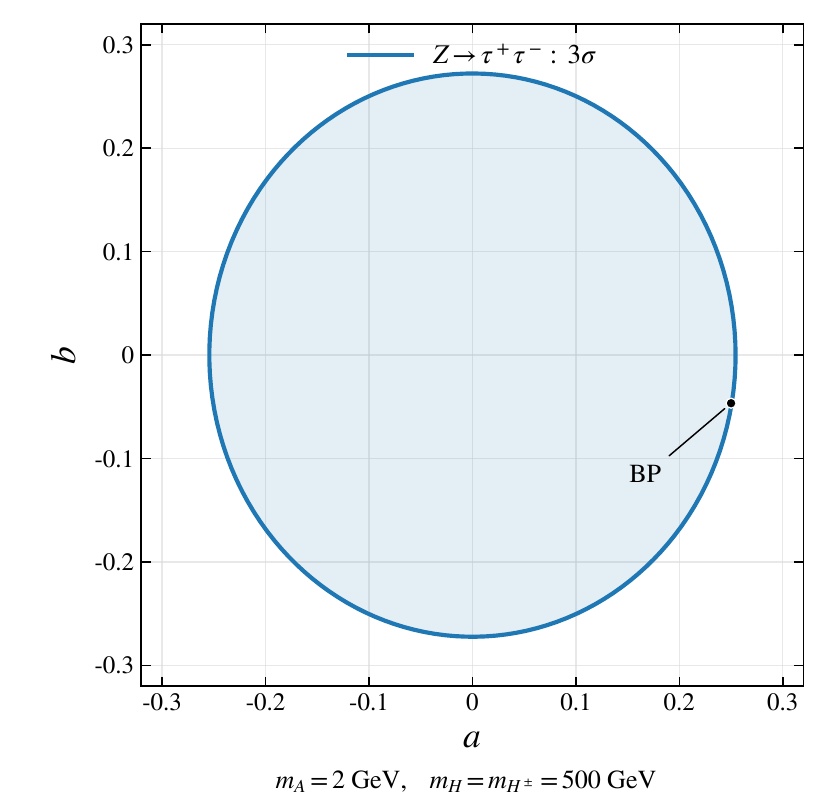}
    \caption{The $3\sigma$ $Z\to\tau^+\tau^-$ compatibility region for $m_A=2$~GeV and $m_H=m_{H^\pm}=500$~GeV. The blue curve is the boundary in Eq.~\eqref{eq:precision_limits}; the shaded interior satisfies that condition. The black point denotes the benchmark in Eq.~\eqref{eq:revised_BP}. This single-ratio comparison uses virtual-width corrections and neglects experimental width correlations; it is not a joint electroweak fit.}
    \label{fig:ab_constraints}
\end{figure}

Among the quantitative comparisons implemented here, lepton universality in $Z\to\tau^+\tau^-$ gives the strongest restriction on $a$ and $b$. We use $R_Z=\Gamma_{\tau\tau}/\Gamma_\ell$, with $\Gamma_\ell=(\Gamma_{ee}+\Gamma_{\mu\mu})/2$, and write $R_Z^{\rm th}=R_Z^{\rm SM}(1+\Delta_Z)$. We include the neutral- and charged-scalar vertex diagrams and on-shell tau external-leg counterterms. Related one-loop $Z$-decay diagram sets and lepton-universality analyses in leptophilic 2HDMs are given in Refs.~\cite{Chun:2016LFU,CiciDag:2026LFU}; the coefficients below are evaluated for our tau-specific texture:
\begin{equation}
 \Delta_Z=-0.07375485\,a^2-0.06438424\,b^2.
 \label{eq:precision_shifts}
\end{equation}
This is the leading virtual two-body width shift. Using the non-universal width inputs of Ref.~\cite{PDG:EW2025} gives $d_Z\equiv R_Z^{\rm exp}/R_Z^{\rm SM}-1=(3.6949\pm2.8224)\times10^{-3}$. The single-ratio requirement $|\Delta_Z-d_Z|\leq3\sigma_Z$ then yields
\begin{equation}
 0.07375485\,a^2+0.06438424\,b^2\leq4.77226\times10^{-3}.
 \label{eq:precision_limits}
\end{equation}
The corresponding ellipse is shown in Fig.~\ref{fig:ab_constraints}. The $3\sigma$ label denotes this one-observable Gaussian compatibility test, not a joint two-parameter confidence region. At $95\%$, its right-hand side would be $1.83689\times10^{-3}$.

We choose a scalar-dominated benchmark just inside this boundary,
\begin{equation}
 a=0.250,\qquad b=-0.04651,
 \label{eq:revised_BP}
\end{equation}
which gives $\Delta_Z\simeq-4.749\times10^{-3}$, a $2.99\sigma$ residual. This point is outside the stricter $95\%$ $Z$ interval; it illustrates the remaining dipole signal near the stated $3\sigma$ boundary. The following secondary checks do not further reduce the displayed region within the approximations used.

For $\tau\to e\bar\nu_e\nu_\tau$, integration of the new-scalar/tree-level interference over three-body phase space, with $m_e=0$ and in the Fermi limit, gives a relative partial-width shift of $-2.006\times10^{-3}$ at the benchmark. The inferred coupling ratio is $g_\tau/g_\mu=0.998997$, within $1.86\sigma$ of the HFLAV value $1.0016\pm0.0014$~\cite{HFLAV:Tau2023}. The individual $95\%$ tau-decay condition contains the entire $Z$ ellipse. Real $A$ emission in tau decay is kinematically forbidden for the chosen mass.

{
The scalar interaction can modify the relation between the tau mass and its Higgs Yukawa coupling~\cite{Stockdale:2025sxi}. Within the $\kappa$ formalism~\cite{LHCHiggsCrossSectionWorkingGroup:2012nn,LHCHiggsCrossSectionWorkingGroup:2013rie}, measurements of $h\to\tau\tau$ by ATLAS and CMS~\cite{ATLAS:2018ynr,CMS:2022dwd} constrain this deviation.
}
Retaining the new-scalar vertices, including $hHH$ and $hH^+H^-$, the tau mass/residue terms and the heavy-scalar Higgs residue, we obtain an interference-order partial-width ratio of $0.9783$ at the benchmark ($m_h=125.25$~GeV), and $0.9712$--$0.9785$ throughout the $Z$ ellipse. With SM Higgs production and total width, this agrees with the adopted signal strength $\mu_{\tau\tau}=0.91\pm0.09$~\cite{ParticleDataGroup:2024cfk}. Within this approximation, Higgs data neither tighten Eq.~\eqref{eq:precision_limits} nor require $|a|>|b|$.

The BESIII radiative $J/\psi\to\gamma A$ search~\cite{BESIII:2022rzz} and OPAL multiphoton data~\cite{OPAL:2002vhf,Knapen:2016moh} constrain the tau-loop-induced CP-even and CP-odd photon couplings, proportional to $a$ and $b$, respectively. Using the conversions of Ref.~\cite{Huang:2025ghw} at the respective photon virtualities, these comparisons are weaker than Eq.~\eqref{eq:precision_limits} at $m_A=2$~GeV and are not plotted separately.

Associated production is a distinct direct-search issue: the benchmark gives the fixed-$\alpha(0)$ Born rate $\sigma(e^+e^-\to\tau^+\tau^-A)\simeq0.691$~pb at $10.58$~GeV and, including off-shell $H$ exchange, $\mathrm{BR}(Z\to\tau^+\tau^-A)\simeq6.55\times10^{-5}$. A tau-loop-dominated $A\to\gamma\gamma$ decay is prompt, but these rates alone do not establish a direct-search exclusion or compatibility.

\begin{table}[h]
    \centering
    \renewcommand{\arraystretch}{1.5} 
    \begin{tabular}{ccc}
        \hline\hline
        $\sqrt{q^2}$ [GeV] & $\text{Re}(a_\tau^{\text{QED}})$ & $\text{Im}(a_\tau^{\text{QED}})$ \\
        \hline
        $3.686$ & $-1.11\times10^{-3}$ & $6.39\times10^{-3}$ \\
        $10.58$ & $-2.44\times10^{-4}$ & $2.18\times10^{-4}$\\
        $0$     & $1.16\times10^{-3}$ & $0$ \\
        \hline\hline
    \end{tabular}
    \caption{Standard Model QED contributions to the $a_\tau$ form factor at the STCF ($3.686$GeV) and Belle II ($10.58$GeV) collision energies, alongside the static limit. These values serve as the baseline to illustrate the interference effects with the NP contributions.}
    \label{tab:QED for a}
\end{table}

\begin{samepage}
Using the quadratic Yukawa dependence, the original rounded dipole predictions are rescaled by $(0.250/0.86)^2\simeq0.08451$. At $\sqrt{s}=3.686$~GeV this gives
\begin{equation}
    a_\tau^{\text{NP}}\big((3.686~\text{GeV})^2\big) \simeq (0.93 + 4.48 i)\times 10^{-4} \,.
    \label{eq:BP_mdm_STCF}
\end{equation}
\end{samepage}
And in the Belle/Belle II center-of-mass energy($\Upsilon(4s)$), we have
\begin{equation}
    a_\tau^{\text{NP}}\big((10.58~\text{GeV})^2\big) \simeq (-0.676 + 1.775i)\times10^{-4} \,.
    \label{eq:BP_mdm_Belle}
\end{equation}
And for the analytical continuation to the $s=0$ point, we can have the quantity that can be measured by ATLAS~\cite{ATLAS:2022ryk},
\begin{equation}
    a_\tau^{\text{NP}}\big(0\big) \simeq 3.55\times10^{-4} \,.
    \label{eq:BP_mdm_static}
\end{equation}
which is consistent with the current limit of ATLAS~\cite{ATLAS:2022ryk} in Eq.~(\ref{Atlas}).

Table~\ref{tab:QED for a} supplies the one-loop QED reference. The NP real part partially offsets QED at $3.686$~GeV but has the same sign as QED at $10.58$~GeV; the imaginary parts add at both energies. The static sum remains within the quoted ATLAS interval.
The EDM is correspondingly reduced to the $10^{-18}\,e\,\mathrm{cm}$ scale at $\sqrt{s}=3.686$~GeV: 
\begin{equation}
    d_\tau^{\text{NP}}\big((3.686~\text{GeV})^2\big) \simeq (1.66+0.930i)\times10^{-18}~e\,\mathrm{cm}\,.
    \label{eq:BP_edm_STCF}
\end{equation}
And in the Belle/Belle II center-of-mass energy, we have
\begin{equation}
    d_\tau^{\text{NP}}\big((10.58~\text{GeV})^2\big) \simeq (-1.42+2.67i)\times10^{-19}~e\,\mathrm{cm}\,.
    \label{eq:BP_edm_Belle}
\end{equation}
which is consistent with the current limit of Belle~\cite{Belle:2021ybo} in Eq.~\ref{Belle} .
For the analytical continuation to the $s=0$ point, we have
\begin{equation}
    d_\tau^{\text{NP}}\big(0\big) \simeq 5.92\times10^{-19}~e\,\mathrm{cm}\,.
    \label{eq:BP_edm_static}
\end{equation}

The pointlike indirect bound in Eq.~\eqref{eq:tradi} is not directly a constraint on this full light-mediator form factor; a model-specific electron-EDM matching calculation is not performed here. For the benchmark in Eq.~\eqref{eq:revised_BP}, $|\mathrm{Im}\,a_\tau^{\rm NP}|/(\Delta\mathrm{Im}\,a_\tau)_{\rm stat}\simeq3.5$ at Belle~II using Table~\ref{Tableg-2}. This is a statistical signal-to-error ratio, not an experimental discovery significance. We next present the extraction strategy independently of this illustrative model point.

\section{Experimental measurements of dipole moments}\label{sec:measurements}

Experimental measurements of $a_\tau$ and $d_\tau$ pose significant challenges due to the tau lepton's rapid decay and the unavoidable presence of missing energy from neutrinos in the final states. Recently, several efforts have been made to improve experimental sensitivities~\cite{Pich:2024qob,Beresford:2024dsc,Gogniat:2025eom,Dittmaier:2025ikh,Buttazzo:2026amk}. Facilities such as Belle II and the STCF offer promising environments for achieving these goals.

We first outline the strategies for extracting $d_\tau$ at low-energy $e^{+}e^{-}$ colliders through the process $e^+ (p_1) e^- (p_2) \to \tau^+ (k_1) \tau^- (k_2)$. Following the notation established in Ref.~\cite{He:2025ewk}, we define 
\begin{eqnarray}
    p_1 = (\sqrt{s}/2,- |\boldsymbol{p}|\hat{\boldsymbol{p}}) ,\quad p_2 = (\sqrt{s}/2, |\boldsymbol{p}|\hat{\boldsymbol{p}}), \nonumber \\
    k_1 = (\sqrt{s}/2,- |\boldsymbol{k}|\hat{\boldsymbol{k}}) ,\quad k_2 = (\sqrt{s}/2, |\boldsymbol{k}|\hat{\boldsymbol{k}}).
\end{eqnarray}
The $\boldsymbol{\hat{l}}_\pm$ denote the momenta of the charged hadrons $h^\pm$ in the $\tau^\pm$ rest frames.

Spin-density-matrix methods for fermion-pair production and decay are also used in the baryon studies of Refs.~\cite{He:2022jjc,Du:2024jfc}; for tau-pair applications we follow Refs.~\cite{Bernreuther:2021elu,He:2025ewk}. As the reconstruction of both $a_\tau$ and $d_\tau$ relies crucially on $\tau$ spin information, we exploit the fact that this information is encoded in the angular distributions of the secondary hadrons produced in the sequential decays $\tau^\pm \to h^\pm \nu$. In the $\tau^\pm$ rest frame, the emission direction of the hadron is correlated with the parent $\tau^\pm$ spin according to
\begin{equation}\frac{1}{\Gamma}\frac{d\Gamma(\tau^\mp\to h^\mp \nu)}{d\cos\theta_\mp}=\frac{1}{2} \left(1\pm \alpha_h \cos\theta_\mp\right).
\end{equation}
where the angular variable is defined as $\cos \theta_\pm \equiv \hat{\boldsymbol{k}} \cdot \boldsymbol{\hat{l}}_\pm$.
In the two-body decay channels, $h^\mp$ represents either $\pi^\mp$ or $\rho^\mp$ mesons. The sensitivity to the $\tau$ spin is governed by the helicity dependent parameters, which are determined to be $(\alpha_\pi, \alpha_\rho) = (1, 0.45)$~\cite{Bernreuther:2021elu,Bernreuther:2021uqm}. Although the $\pi^\mp$ mode serves as an ideal spin analyzer due to its maximal sensitivity ($\alpha_\pi = 1$), the $\rho^\mp$ channel possesses a higher branching ratio. Consequently, the higher statistics provided by the $\rho^\mp$ mode compensate for its lower sensitivity, leading to a statistical weight comparable to the $\pi^\mp$ mode.

The operators for $d_\tau$ measurement in Ref.~\cite{He:2025ewk} are given as 
\begin{eqnarray}
    \mathcal{O}_1 = \hat{\boldsymbol{l}}_- \cdot \hat{\boldsymbol{k}}, \quad
    \mathcal{O}_2 =\hat{\boldsymbol{l}}_+ \cdot \hat{\boldsymbol{k}},\quad
    \mathcal{O}_3 = (\hat{\boldsymbol{l}}_- \times \hat{\boldsymbol{l}}_+) \cdot \hat{\boldsymbol{k}}, 
    \quad 
    \mathcal{O}_4 =(\hat{\boldsymbol{p}} \cdot \hat{\boldsymbol{k}}) \big[ (\hat{\boldsymbol{l}}_- \times \hat{\boldsymbol{l}}_+) \cdot \hat{\boldsymbol{p}} \big].
\end{eqnarray}
Measuring the average values $\braket{\mathcal{O}_i}$, the real and imaginary part of $d_\tau$ can be isolated and are given by 
\begin{eqnarray}
   \mathrm{Im}\left( d_\tau \right) &=& \frac{-e(3s + 6m_\tau^2)}{4m_\tau\sqrt{s}\sqrt{s - 4m_\tau^2}} \left( \frac{\langle \hat{\boldsymbol{l}}_- \cdot \hat{\boldsymbol{k}} \rangle}{\alpha_h} + \frac{\langle \hat{\boldsymbol{l}}_+ \cdot \hat{\boldsymbol{k}} \rangle}{\bar{\alpha}_{h'}} \right),\\
   \mathrm{Re}\left( d_\tau \right)^a &=& e \frac{9}{4} \frac{s + 2m_\tau^2}{\alpha_h \bar{\alpha}_{h'} m_\tau \sqrt{s^2 - 4sm_\tau^2}} \langle (\hat{\boldsymbol{l}}_- \times \hat{\boldsymbol{l}}_+) \cdot \hat{\boldsymbol{k}} \rangle,\\
    \mathrm{Re}\left( d_\tau \right)^b &=& e \frac{45}{2} \frac{(s + 2m_\tau^2) \langle (\hat{\boldsymbol{p}} \cdot \hat{\boldsymbol{k}}) (\hat{\boldsymbol{l}}_- \times \hat{\boldsymbol{l}}_+) \cdot \hat{\boldsymbol{p}} \rangle}{\alpha_h \bar{\alpha}_{h'} \sqrt{s}(\sqrt{s} - 2m_\tau) \sqrt{s - 4m_\tau^2}}.
\end{eqnarray}
where the superscripts $a)$ and $b)$ distinguish between two measurement methods. As demonstrated in Ref.~\cite{He:2025ewk}, $\mathrm{Im}\left( d_\tau \right)$ can be extracted using only the laboratory frame energies of the final-state hadrons. In contrast, measuring the real part is more involved, as it requires the full reconstruction of the $\tau$ momentum. To resolve the kinematic ambiguities introduced by undetected neutrinos, Ref.~\cite{He:2025ewk} suggested a selection technique that utilizes events where the $\tau$ travel distance exceeds the detector's spatial resolution. They conclude that, at the STCF, the precision for both 
   \(\mathrm{Re}(d_\tau)\) and 
	\(\mathrm{Im}(d_\tau)\) reaches its peak at a center-of-mass energy of \(6.3\,\mathrm{GeV}\), with attainable sensitivities of  
{ 	$
2.8  \times 10^{-18}\,e\,\mathrm{cm}$   and $0.7 \times 10^{-18}\,e\,\mathrm{cm},
	$ } 
	respectively.

In the following, we adapt similar methods to enhance the sensitivity for $a_\tau$ measurements within the same process and techniques.
The operators we choose to achieve the goal are
 \begin{eqnarray}
     \mathcal{O}_5 =\left(\hat{\boldsymbol{p}} \cdot \hat{\boldsymbol{k}}\right)^2, \quad
    \mathcal{O}_6 = \frac{(\hat{\boldsymbol{p}} \cdot \hat{\boldsymbol{k}}) \big[ \hat{\boldsymbol{l}}_\mp \cdot (\hat{\boldsymbol{p}} \times \hat{\boldsymbol{k}}) \big]}{|\hat{\boldsymbol{p}} \times \hat{\boldsymbol{k}}|}.
 \end{eqnarray}
The real and imaginary $a_\tau$ are given by
\begin{eqnarray}
\mathrm{Re}\left( a_\tau \right) &=& \frac{2(s+2m_\tau^2)(s+m_\tau^2)}{s (s-4m_\tau^2)} - \frac{5 (s+2m_\tau^2)^2}{s(s-4m_\tau^2)}\left\langle  \left(\hatp \cdot \hatk \right)^2\right\rangle,   \\
\mathrm{Im}\left( a_\tau \right) &=& - \frac{64 m_\tau (2 m_\tau^2 +s)}{\pi \alpha_{h^\pm} \sqrt{s} (s-4m_\tau^2)} \left\langle { \frac{(\hatp \cdot \hatk) \left[\hatl_\mp \cdot (\hatp\times \hatk)\right]}{|\hatp\times\hatk|}}\right\rangle.
\end{eqnarray}

We retain linear dipole contributions and use the single-photon production approximation. The following estimates quantify statistical uncertainties only; they are not total experimental sensitivities. For a normalized sample moment $\langle\mathcal O_i\rangle=N^{-1}\sum_j\mathcal O_i^{(j)}$, a common cross-section or luminosity normalization cancels, but angular acceptance, resolution, charge-dependent response, backgrounds, and radiative distortions need not cancel. We do not estimate these systematics. The statistical errors are
\begin{eqnarray}
\Delta\mathcal{\hat{O}}_i = \sqrt{\frac{\langle\mathcal{\hat{O}}_i^2\rangle-\mathcal{\langle{\hat{O}}}_i\rangle^2 }{N}}, \quad i =1,2,...,6, 
\end{eqnarray}
resulting in 
\begin{eqnarray}
 \Delta \mathrm{Im}\left( d_\tau \right) &=& \frac{e(s + 2m_{\tau}^2)}{4m_{\tau}\sqrt{s}\sqrt{s - 4m_{\tau}^2}} \sqrt{ \frac{6}{ N_{\mathrm{eff}}}  }, \\
    \Delta \mathrm{Re}\left( d_\tau \right)(D)^a &=& \frac{3 e}{4}
		\frac{
			s + 2 m_\tau^2 
		}{
			m_\tau  
			\sqrt{
				s^2 - 4 s m_\tau^2 
			} 
		}
		\sqrt{ 
			\frac{2}{N'_{\text{eff}}}
		}
		\,,\\
        \Delta \mathrm{Re}\left( d_\tau \right)(D)^b &=& \frac{3e}{2} \frac{\sqrt{s^2 + 3sm_{\tau}^2 + 2m_{\tau}^4}}{\sqrt{s}(\sqrt{s} - 2m_{\tau})\sqrt{s - 4m_{\tau}^2}} \sqrt{\frac{20}{N'_{\mathrm{eff}}}}.
\end{eqnarray}
and  
\begin{eqnarray}
    \Delta \mathrm{Re}\left( a_\tau \right) &=& \frac{(s+2m_\tau^2)}{s(s-4m_\tau^2)} \sqrt{\frac{32 m_\tau^4+64 m_\tau^2 s+17 s^2}{14N_\mathrm{eff}^{(1)}}} ,\\
    \Delta \mathrm{Im}\left( a_\tau \right) &=& \frac{64 m_\tau}{\pi  (s-4m_\tau^2)\sqrt{s}} \sqrt{
    \frac{4m_\tau^4 + 6m_\tau^2s+2s^2}{30 N_{\mathrm{eff}}^{(2)}}
    }.
\end{eqnarray}
Here $N_{\rm eff} = \epsilon
N_{\tau^+\tau^-}
\left(
\alpha_\pi^2 	{\rm Br}_\pi + \alpha_\rho^2 {\rm Br}_\rho
\right)$, $N'_{\rm eff} =\epsilon P_\tau
N_{\tau^+\tau^-}
\left(
\alpha_\pi^2 	{\rm Br}_\pi + \alpha_\rho^2 {\rm Br}_\rho
\right)^2$, $N_\mathrm{eff}^{(1)} = \epsilon N_{\tau^+\tau^-} P_\tau \left(\mathrm{Br}_\rho + \mathrm{Br}_\pi\right)$, and $N_{\rm eff}^{(2)} =\epsilon P_\tau
N_{\tau^+\tau^-}
\left(
\alpha_\pi^2 	{\rm Br}_\pi + \alpha_\rho^2 {\rm Br}_\rho
\right)$. $\epsilon$ is the selection signal efficiency, $P_\tau$ is the probability of detecting the $\tau$'s direction and $\mathrm{Br}_{\pi(\rho)}$ is the branching ratio of the process $\tau^- \to \pi (\rho)+ \nu_\tau$. We take $\epsilon=3\%$ as an illustrative effective efficiency, not a detector-derived value. The simulation study of Ref.~\cite{Sun:2024vcd} includes event selection, backgrounds and tau reconstruction in the $\rho\rho$ channel, with $6.3\%$ signal efficiency and $80.0\%$ purity. It illustrates the analysis dependence of experimental performance, not that systematics are negligible for the observables used here.

While Ref.~\cite{He:2025ewk} primarily examined $d_\tau$ performance at the STCF, we extend that analysis to Belle II using the same operators. For $e^+e^-\to\tau^+\tau^-$ at $\sqrt{s}=10.58$~GeV with $N_{\tau^+\tau^-} \sim 4.5\times 10^{10}$ events~\cite{Belle-II:2018jsg,Belle-II:2022cgf}, the resulting sensitivities are $\Delta {\rm Im} \left( d_\tau \right) \approx 5.19 \times 10^{-19} e\,\mathrm{cm}$ and $\Delta {\rm Re}\left( d_\tau \right) \approx 7.62 \times 10^{-19} e\,\mathrm{cm}$. 

\begin{figure}[htbp]
    \centering
    \includegraphics[width=0.48\textwidth]{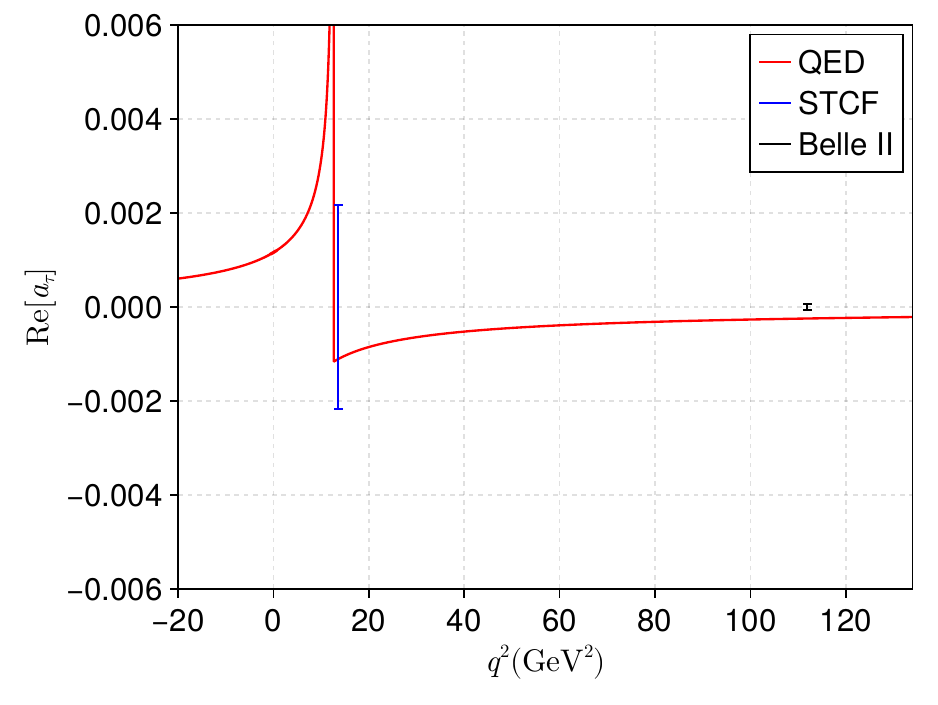}
    \hfill
    \includegraphics[width=0.48\textwidth]{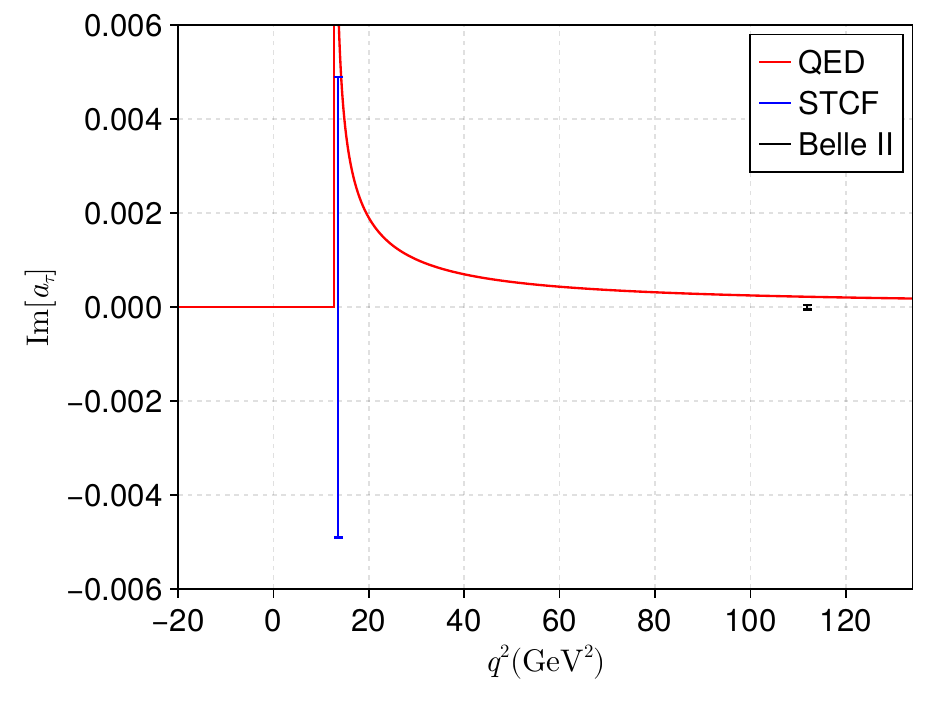}
    \caption{One-loop QED predictions for $\mathrm{Re}\left( a_\tau \right)(q^2)$ and $\mathrm{Im}\left( a_\tau \right)(q^2)$, alongside projected statistical-only $1\sigma$ sensitivities at the $\psi(2S)$ and $\Upsilon(4S)$ resonances. Notably, the discontinuities observed in the distributions arise at the threshold $q^2 = 4m_\tau^2$. }
    \label{fig:g-2}
\end{figure}

We retain the $\psi(2S)$ and $\Upsilon(4S)$ energies as reference points for the statistical comparison with one-loop QED in Table~\ref{Tableg-2} and Fig.~\ref{fig:g-2}. Two-photon exchange and other radiative contributions to the MDM observables are not evaluated here, and no quantitative suppression at the chosen energies is assumed. CP-odd EDM observables suppress CP-even backgrounds for ideal symmetric acceptance, but detector asymmetries and imperfect subtraction can still induce biases. A realistic forecast requires their nuisance-parameter treatment in addition to the statistical errors above.

\begin{table}[h]
    \centering
    \caption{Projected statistical-only sensitivities to $g-2$ at $\sqrt{s}=m_{\psi(2s)}$ and $\sqrt{s} = m_{\Upsilon(4S)}$. The one-loop QED prediction are also shown in this table.  Here we consider both $\pi$ and $\rho$ decay channel, and take the spatial resolution $D=30 \mathrm{\mu m}$ and $15 \mathrm{\mu m}$ for STCF and Belle II, respectively, as an illustration. At STCF, the luminosity is expected to reach $1~ab^{-1}$ per year. Over ten years of data collection, the total number of $\psi(2S)$ events is anticipated to be about $2 \times 10^{10}$~\cite{Achasov:2023gey}. At Belle II, the events number $N_{\tau^+\tau^-} \approx 4.5 \times 10^{10}$~\cite{Belle-II:2018jsg,Belle-II:2022cgf}.}
    \label{Tableg-2}
    \begin{tabular}{l c c c c}
        \hline\hline
        \multicolumn{1}{c}{$\sqrt{q^2}$ [GeV]} & $|\mathrm{Re} (a_\tau^{QED})|$ &
        $\Delta\mathrm{Re}\left( a_\tau \right)$  & $|\mathrm{Im}(a_\tau^{QED})|$ &
        $\Delta\mathrm{Im}\left( a_\tau \right)$  \\
        \hline
        $m_{\psi(2S)}$(3.686GeV) & $1.11\times 10^{-3}$ &
        {$2.17\times 10^{-3}$} & $6.39\times 10^{-3}$  &
        {$4.90\times 10^{-3}$} \\
        {$m_{\Upsilon(4S)}$(10.58GeV)} & 
        $2.44\times 10^{-4}$ &
        $6.29\times 10^{-5}$&
        $2.18\times10^{-4}$&
        $5.02\times 10^{-5}$
         \\
        \hline\hline
    \end{tabular}
\end{table}

The angular and spin-correlation observables provide separate access to the real and imaginary form factors. Under the stated event-count and reconstruction assumptions, the statistical errors approach the one-loop QED scale, especially for the imaginary part and at Belle~II. The two energies offer complementary kinematic information. Realizing this reach, or converting the benchmark signal-to-error ratio into a significance, requires control of the unquantified systematic effects.

\section{Conclusion}

In this work, we have investigated the $q^2$ dependence and absorptive imaginary parts of the tau MDM and EDM. The heavy-scale LEFT/SMEFT illustration and the explicit light-scalar 2HDM calculation are separate: the light state is retained in the latter, and no EFT with a cutoff comparable to the collision energy is used as a numerical benchmark. Local dipoles and four-fermion interactions contribute differently to timelike form factors; a general assertion that the latter evade all electron-EDM constraints is not made.

Within the tau-specific 2HDM, precision $Z$ and electronic tau decays substantially reduce the original Yukawa benchmark. At $m_A=2$~GeV and $m_H=m_{H^\pm}=500$~GeV, we use $(a,b)=(0.250,-0.04651)$ near the adopted $3\sigma$ $Z$ boundary and inside the individual $95\%$ tau-decay interval. The resulting imaginary MDM at Belle~II is approximately $1.78\times10^{-4}$, while the EDM is of order $10^{-18}\,e\,\mathrm{cm}$ at $3.686$~GeV.The BP provides a representative light-scalar scenario consistent with the precision-decay constraints adopted in this work and serves to illustrate the experimental sensitivity to the complex tau dipole form factors and their $q^2$ dependence.

The proposed angular and spin observables separately probe the real and imaginary dipole form factors. Their statistical-only projections motivate complementary measurements at Belle~II and STCF and the study of momentum dependence. The near-boundary BP yields a Belle~II imaginary-MDM signal about $3.5$ times the projected statistical error; this does not include systematic uncertainties or establish a discovery forecast. Detector response, backgrounds and radiative corrections must be controlled in an experimental implementation.

\begin{acknowledgments}
This work is supported in part by the National Key Research and Development Program of China under Grant
No. 2020YFC2201501, by the Fundamental Research
Funds for the Central Universities, by National Natural Science Foundation of P.R. China (No.12090064, 12375088, 12575096 and W2441004).
\end{acknowledgments}

\appendix
\section{Scalar-sector conventions in the two-Higgs-doublet model}
\label{2HDM}
In this appendix, we show how the effective scalar interactions used in the main text can arise from a renormalizable 2HDM, where the two scalar doublets $\Phi_{1,2}$ transform under the SM gauge group $SU(3)_C\times SU(2)_L\times U(1)_Y$ as $(1,2,1/2)$. The most general scalar potential can be written as~\cite{Wu:1994ja}
\begin{eqnarray}	\begin{aligned}V&=\frac{1}{2}\left[m_{11}^{2}|\Phi_{1}|^{2}+m_{22}^{2}|\Phi_{2}|^{2}-\left(m_{12}^{2}\Phi_{1}^{\dagger}\Phi_{2}+\mathrm{h.c.}\right)\right]+\frac{\lambda_{1}}{2}(\Phi_{1}^{\dagger}\Phi_{1})^{2}+\frac{\lambda_{2}}{2}(\Phi_{2}^{\dagger}\Phi_{2})^{2}\\&+\lambda_{3}(\Phi_{1}^{\dagger}\Phi_{1})(\Phi_{2}^{\dagger}\Phi_{2})+\lambda_{4}(\Phi_{1}^{\dagger}\Phi_{2})(\Phi_{2}^{\dagger}\Phi_{1})+\left[\frac{\lambda_{5}}{2}(\Phi_{1}^{\dagger}\Phi_{2})^{2}+\left(\lambda_6\Phi_1^\dagger\Phi_1+\lambda_7\Phi_2^\dagger\Phi_2\right)\left(\Phi_1^\dagger\Phi_2\right)+\mathrm{h.c.}\right].\end{aligned}
\label{general potential}
\end{eqnarray}
Consequently, the parameters $m_{12}^2$, $\lambda_5$, $\lambda_6$, and $\lambda_7$ are in general allowed to be complex~\cite{Carena:2014nza,BhupalDev:2017txh}. The scalar doublets are parameterized as~\cite{Inoue:2014nva,Chun:2019oix}
\begin{eqnarray}
\Phi_1=\begin{pmatrix}\phi_1^+\\\frac{1}{\sqrt{2}}(v_1+\phi_1^{0r}+i\phi_1^{0i})\end{pmatrix},\quad\Phi_2=\begin{pmatrix}\phi_2^+\\\frac{1}{\sqrt{2}}(v_2+\phi_2^{0r}+i\phi_2^{0i})\end{pmatrix}.
\end{eqnarray}
where $v^2\equiv v_1^2+v_2^2=(246\,\mathrm{GeV})^2$, with $\tan\beta\equiv v_2/v_1$.

For the neutral Higgs sector
\begin{eqnarray}
\begin{pmatrix}G^0\\A^0\end{pmatrix}=\begin{pmatrix}\cos\beta&\sin\beta\\-\sin\beta&\cos\beta\end{pmatrix}\begin{pmatrix}\phi_1^{0i}\\\phi_2^{0i}\end{pmatrix}.
\end{eqnarray}
An orthogonal rotation matrix $R$ is defined to diagonalize the neutral mass matrix by $R\mathcal{M}^2R^\mathrm{T}=\mathrm{diag}(m_{h_1}^2,m_{h_2}^2,m_{h_3}^2)$, where
\begin{eqnarray}
	\mathcal{L}_{mass}=\frac{1}{2}\left(h_1h_2h_3\right)\mathcal{M}_{diag}^2\begin{pmatrix}h_1\\h_2\\h_3\end{pmatrix},\mathrm{~where~}\begin{pmatrix}h_1\\h_2\\h_3\end{pmatrix}=R\begin{pmatrix}\phi_1^{0r}\\\phi_2^{0r}\\A^0\end{pmatrix}.
\end{eqnarray}
and
\begin{eqnarray}
	R=\begin{pmatrix}-s_\alpha c_{\alpha_b}&c_\alpha c_{\alpha_b}&s_{\alpha_b}\\s_\alpha s_{\alpha_b}s_{\alpha_c}-c_\alpha c_{\alpha_c}&-s_\alpha c_{\alpha_c}-c_\alpha s_{\alpha_b}s_{\alpha_c}&c_{\alpha_b}s_{\alpha_c}\\s_\alpha s_{\alpha_b}c_{\alpha_c}+c_\alpha s_{\alpha_c}&s_\alpha s_{\alpha_c}-c_\alpha s_{\alpha_b}c_{\alpha_c}&c_{\alpha_b}c_{\alpha_c}\end{pmatrix}.
\end{eqnarray}
If we assume there is no symmetry in the Yukawa couplings, as
$	-\mathcal{L}_Y=Y_1\bar{L_L}\Phi_1\ell_R+Y_2\bar{L_L}\Phi_2\ell_R+h.c.,$
so the mass of lepton can be written as $m_l=(Y_1 v\cos\beta)/\sqrt{2}+(Y_2v\sin\beta)/\sqrt{2}$. Then the Yukawa interaction can be rewritten by
\begin{eqnarray}
	\begin{aligned}
		\mathcal{L}_{\mathrm{Yukawa}}^h=&-\sum_{i=1}^3\left[ \left( \frac{m_f}{v \sin \beta}R_{i2}+(\frac{Y_1 (R_{i1}- \cot \beta R_{i2})}{\sqrt{2}}   ) \right)(h_i\overline{f}f)+i ( \frac{m_f \cos \beta}{v \sin\beta}  -\frac{Y_1 }{\sqrt{2}\sin\beta}    )R_{i3} (h_i\overline{f}f)\gamma_5\right].
	\end{aligned}
\end{eqnarray}

For simplicity, we can also work in the Higgs basis to show the result. 
We can do the transformation from the basis we discussed before to Higgs basis~\cite{Gunion:2002zf,Davidson:2005cw}, where the vev only form the first the Higgs doublet. The structure of $V(\Phi_{1,2})$ is preserved under the rotation  
\begin{eqnarray}
\begin{pmatrix}\Phi_1'\\\Phi_2'\end{pmatrix} =
\begin{pmatrix}
\cos\beta & \sin\beta \\
-\sin\beta & \cos\beta
\end{pmatrix}
\begin{pmatrix}\Phi_1\\\Phi_2\end{pmatrix}\,.
\end{eqnarray}
with the parameters in the scalar potential transforming linearly as well.
In the new basis, the scalar potential in the Higgs basis can be written by
\begin{eqnarray}
\begin{aligned}V(\Phi_1^\prime,\Phi_2^\prime)&\begin{aligned}=M_{11}^2\Phi_1^{\prime\dagger}\Phi_1^\prime+M_{22}^2\Phi_2^{\prime\dagger}\Phi_2^\prime-\left(M_{12}^2\Phi_1^{\prime\dagger}\Phi_2^\prime+\mathrm{c.c.}\right)\end{aligned}\\&+\frac{1}{2}\Lambda_1\left(\Phi_1^{\prime\dagger}\Phi_1^\prime\right)^2+\frac{1}{2}\Lambda_2\left(\Phi_2^{\prime\dagger}\Phi_2^\prime\right)^2+\Lambda_3\left(\Phi_1^{\prime\dagger}\Phi_1^\prime\right)\left(\Phi_2^{\prime\dagger}\Phi_2^\prime\right)+\Lambda_4\left(\Phi_1^{\prime\dagger}\Phi_2^\prime\right)\left(\Phi_2^{\prime\dagger}\Phi_1^\prime\right)\\&+\left(\frac{1}{2}\Lambda_5\left(\Phi_1^{\prime\dagger}\Phi_2^\prime\right)^2+\left(\Lambda_6\Phi_1^{\prime\dagger}\Phi_1^\prime+\Lambda_7\Phi_2^{\prime\dagger}\Phi_2^\prime\right)\left(\Phi_1^{\prime\dagger}\Phi_2^\prime\right)+\mathrm{c.c.}\right).
\end{aligned}
\end{eqnarray}
where $v_1^\prime = v$ and $v_2^\prime = 0$, resulting in 
\begin{eqnarray}
\Phi_1^\prime=\begin{pmatrix}G^+\\\frac{1}{\sqrt{2}}\left(v+h+iG^0\right)\end{pmatrix},\quad \Phi_2^\prime=\begin{pmatrix}H^+\\\frac{1}{\sqrt{2}}\left(H+iA\right)\end{pmatrix}.
\end{eqnarray}
So the neutral mass matrix in the basis can be written by 
\begin{eqnarray}
	\frac{M^2}{v^2}=\begin{pmatrix}\Lambda_1&\mathrm{Re}[\Lambda_6]&-\mathrm{Im}[\Lambda_6]\\\\\mathrm{Re}[\Lambda_6]&\frac{M_{H^\pm}^2}{v^2}+\frac{1}{2}(\Lambda_4+\mathrm{Re}[\Lambda_5])&-\frac{1}{2}\mathrm{Im}[\Lambda_5]\\\\-\mathrm{Im}[\Lambda_6]&-\frac{1}{2}\mathrm{Im}[\Lambda_5]&\frac{M_{H^\pm}^2}{v^2}+\frac{1}{2}(\Lambda_4-\mathrm{Re}[\Lambda_5])\end{pmatrix}.
\end{eqnarray}
For $\Lambda_6=0$, the SM-like Higgs field $h$ does not mix with the non-SM neutral scalars, which is called the alignment limit. So
\begin{eqnarray}
M_{H,A}^2=\begin{pmatrix}M_{22}^2+\frac{1}{2}(\Lambda_3+\Lambda_4+\mathrm{Re}[\Lambda_5])v^2&-\frac{1}{2}\mathrm{Im}[\Lambda_5]v^2\\-\frac{1}{2}\mathrm{Im}[\Lambda_5]v^2&M_{22}^2+\frac{1}{2}(\Lambda_3+\Lambda_4-\mathrm{Re}[\Lambda_5])v^2\end{pmatrix}.
\end{eqnarray}
with $m_{h}^{2}=\Lambda_1v^2,m_{H^\pm}^2=M_{22}^2+\frac{1}{2}\Lambda_3v^2$. 
In particular, $m_A^2 = m_{H^\pm}^2 + \frac{1}{2}(\Lambda_4-\Lambda_5)v^2$ in our work. 

The phenomenological assumptions and precision comparisons are summarized in Sec.~\ref{sec:uvmodel}. For the real-potential benchmark, $m_H=m_{H^\pm}$ implies $\Lambda_4=-\Lambda_5$. Setting the renormalized combination $\Lambda_3+\Lambda_4-\Lambda_5$ to zero suppresses the tree-level $hAA$ coupling without eliminating $hHH$ or $hH^+H^-$.

The scalar-potential parameters are subject to the perturbative criterion $|\Lambda_i|<4\pi$ and the bounded-from-below conditions~\cite{Gunion:2002zf}
\begin{eqnarray}
\begin{aligned}\Lambda_{1,2}>0,\quad\Lambda_3>-(\Lambda_1\Lambda_2)^{1/2},\quad\Lambda_3+\Lambda_4-|\Lambda_5|>-(\Lambda_1\Lambda_2)^{1/2}.\end{aligned}
\end{eqnarray}
The mass relation $m_H^2-m_A^2=\Lambda_5v^2$ consequently implies the necessary bound
\begin{eqnarray}
    m_H<\sqrt{m_A^2+|\Lambda_5|_{\max}v^2}.
\end{eqnarray}
These relations allow a light $A$ while the heavier states are degenerate; they do not replace the phenomenological tests discussed in the main text.

\bibliographystyle{utphys}
\bibliography{ref}

\end{document}